\documentclass[11pt,twoside]{article}

\usepackage{asp2004}
\usepackage{epsf}
\usepackage{lscape}

\usepackage{amsmath}
\usepackage{amssymb}

\begin{document}
\title{NLTE wind models for SMC stars}
\author{Ji\v{r}\'{i} Krti\v{c}ka}   
\affil{Masarykova univerzita,
        CZ-611 37 Brno, Czech Republic, \\email:
        krticka@physics.muni.cz}    

\begin{abstract} 
We study stellar wind properties of selected late O stars in the Small
Magellanic Cloud (SMC). We calculate NLTE line-driven wind models for these
stars and compare predicted wind parameters with
observed values. We found satisfactory agreement between theoretical
and observed terminal velocities. On the other hand, predicted and observed
mass-loss rates are in a good agreement only for higher mass-loss rates. For
mass-loss rates lower than approximately
${10^{-7}\,\text{M}_\odot\,\text{year}^{-1}}$ we found large
discrepancy between
theoretical and observed values. We propose a new explanation of this effect
based on
dynamical decoupling of some atoms. Finally, we study the dependence of wind
terminal velocities and mass-loss rates on metallicity.
\end{abstract}



\section{Introduction}   

Hot stars wind properties depend on the
stellar metallicity. Low-metallicity environment of SMC offers a
possibility to study this dependence.
For this purpose we apply wind models of
\citet{nltei}. 


\newcommand\mdot{\dot M}
\newcommand\smrok{\text{M}_{\odot}\,\mathrm{yr}^{-1}}
\newcommand{\dmdt}{\mdot}

Wind parameters of SMC stars predicted by our code were compared with that
derived from observations by \citet{pulmoc}, \citet{bourak}, \citet{maso} and 
\citet{martin}. 

\section{Comparison of observed wind parameters}



\begin{figure}[htb]
\resizebox{0.55\hsize}{!}{\includegraphics{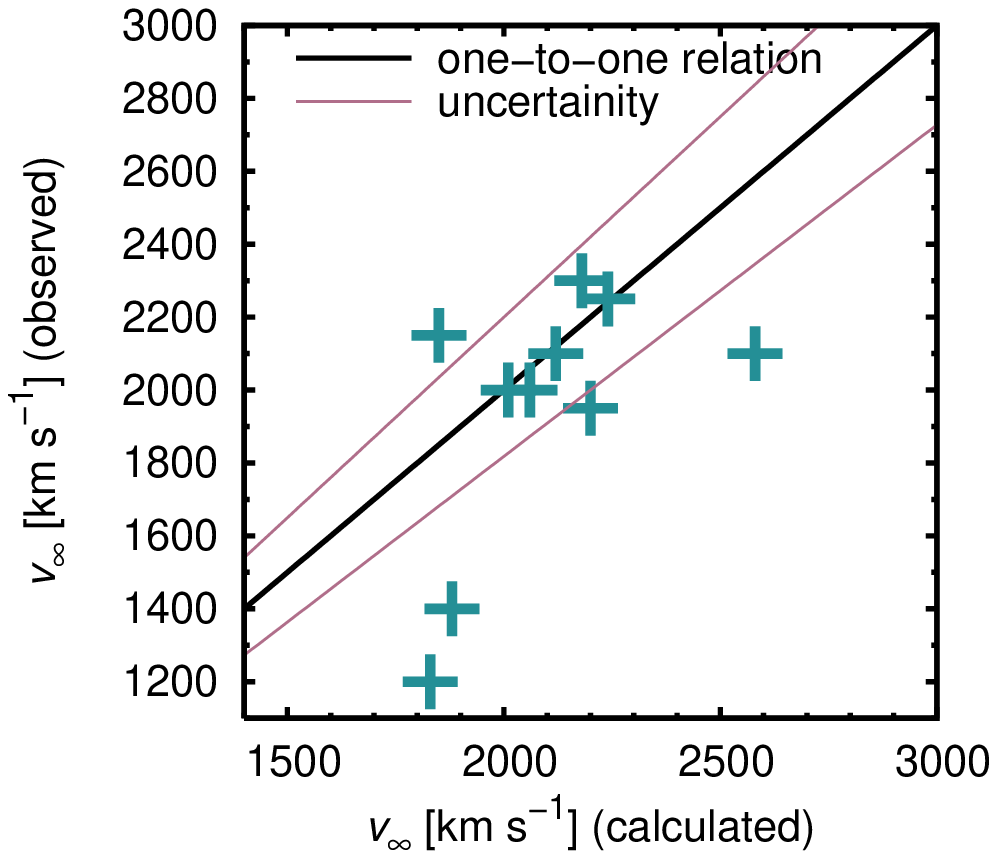}}
\resizebox{0.455\hsize}{!}{\includegraphics{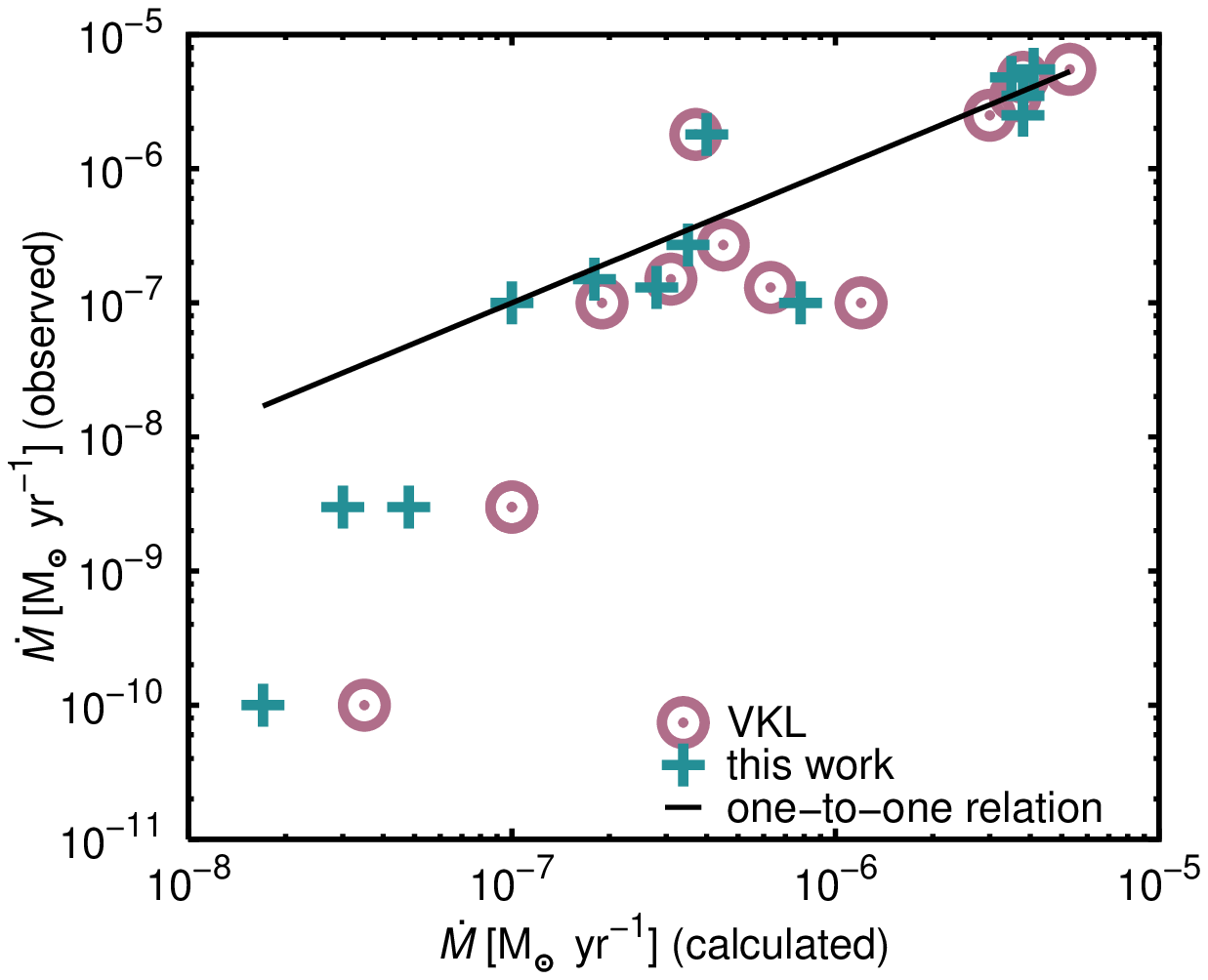}}
\caption{{\em Left:} Comparison of observed and predicted terminal velocities
{\em Right:} Comparison predicted mass-loss rates by us and by \citet[]
[hereafter VKL]{vikolamet} and mass-loss rates derived
from observation} 
\label{poro}
\end{figure}

There is a relatively good agreement between predicted and observed terminal 
velocities (see Fig.~\ref{poro}) with few exceptions.
The mean value of ${v_\infty/v_\text{esc}\sim2.3}$ is slightly
lower than that
of Galactic O stars.
This may indicate that SMC terminal velocities are slightly
lower than that of Galactic stars.



There is a good agreement between calculated and observed mass-loss rates
for stars with high mass-loss rates (${\dmdt
\gtrsim10^{-7}\,\smrok}$). 
However, there is a significant
disagreement between theoretical and observed values for stars with low mass-loss
rates (${\dmdt \lesssim10^{-7}\,\smrok}$). In this case the predicted
mass-loss rates are more than ten times higher than the observed ones
\citep{bourak}.

\section{Multicomponent effects}

Hot star winds have a multicomponent nature which for
low-density or low-metallicity flow influences the wind structure
\citep[e.g.][]{kkii}.
To test the importance of multicomponent effects 
we calculated models with wind components sulphur, remaining
metals, H-He component and electrons.

For stars with relatively good agreement between observed and predicted
mass-loss rates the multicomponent structure can be neglected.
On the other hand, for stars with systematically too high predicted mass-loss
rates either the decoupling of wind components 
occurs or the frictional heating is important.
In this case the decoupling instability may 
lower the mass-loss rate. This may be an explanation of
too high predicted mass-loss rates. However, only for one star studied the
decoupling occurs for velocities lower than the terminal velocity. 
Improved model treatment may help to provide more
consistent explanation.

\section{Variations of wind parameters with metallicity}

To study the variations of wind parameters with metallicity we recalculated wind
models with the same stellar parameters, however
with metallicity ${1.5}$ times higher. 
For higher metallicity the mass loss rate is higher,
\begin{equation}
{\dmdt\sim Z^{0.70}.}
\end{equation}
The terminal velocity varies with metallicity only slightly, on average
${v_\infty\sim Z^{0.05}}$.

\acknowledgements 
This work was supported by grants GA\,\v{C}R 205/03/ D020, 205/04/1267 and 
by the Czech-Estonian academic interchange program.


\begin{thebibliography}{}
\bibitem[Bouret et al.(2003)]{bourak}  Bouret, J.-C. , Lanz, T., Hillier, D. J.
        et al., 2003, ApJ, 595 1182 
\bibitem[Krti\v{c}ka \& Kub\' at(2001)]{kkii} Krti\v{c}ka, J., \& Kub\' at, J.,
	2001, A\&A, 377, 175
\bibitem[Krti\v{c}ka \& Kub\' at(2004)]{nltei} Krti\v{c}ka, J., \& Kub\' at, J.,
	2004, A\&A, 417, 1003
\bibitem[Martins et al.(2004)]{martin} Martins, F. , Schaerer, D., Hillier,
	D. J.
	et al., 2004, A\&A, 420 1087
\bibitem[Massey et al.(2004)]{maso} Massey, P., Bresolin, F., Kudritzki, R. P.
        et al., 2004, ApJ, 608 1001
\bibitem[Puls et al.(1996)]{pulmoc} Puls, J., Kudritzki, R.-P., Herrero, A.,
        et al., 1996, A\&A, 305 171
\bibitem[Vink et al.(2001)]{vikolamet} Vink, J. S., de Koter, A.,
	Lamers, H. J. G. L. M., 2001, A\&A 369, 574
\end{thebibliography}
\end{document}